\documentclass[aps,preprint,superscriptaddress]{revtex4}
\usepackage[dvips]{graphicx,epsfig}
\usepackage{epsfig,amsmath,amsfonts,amssymb,wasysym,color}

\def\be{\begin{equation}}
\def\ee{\end{equation}}\def\la{\langle}
\def\ra{\rangle}

\def\Cas{\mathrm{C}}
\def\bulk{\mathrm{bulk}}

\def\ex{\mathrm{ex}}
\def\kB{k_{\mathrm{B}}}

\def\C{{\mathcal C}}

\def\rme{{\rm e}}
\def\rmd{{\rm d}}
\newcommand\reff[1]{(\ref{#1})}
\def\bea{\begin{eqnarray}}
\def\eea{\end{eqnarray}}

\def\sf{\vartheta}
\def\sfh{\hat\vartheta}
\def\sfb{\theta}
\def\sfhb{\hat\theta}

\begin{document}

\title{%
Monte Carlo simulation results for critical Casimir forces%
}

\author{O.~Vasilyev}
\affiliation{\small Max-Planck-Institut f\"ur Metallforschung,
  Heisenbergstr.~3, 70569 Stuttgart, Germany}
\affiliation{\small Institut f\"ur Theoretische und Angewandte Physik,
Universit\"at Stuttgart, Pfaffenwaldring~57, D-70569 Stuttgart, Germany.}
\author{A.~Gambassi}
\affiliation{\small Max-Planck-Institut f\"ur Metallforschung,
  Heisenbergstr.~3, 70569 Stuttgart, Germany}
\affiliation{\small Institut f\"ur Theoretische und Angewandte Physik,
Universit\"at Stuttgart, Pfaffenwaldring~57, D-70569 Stuttgart, Germany.}
\author{A.~Macio\l ek}
\affiliation{\small Max-Planck-Institut f\"ur Metallforschung,
  Heisenbergstr.~3, 70569 Stuttgart, Germany}
\affiliation{\small Institut f\"ur Theoretische und Angewandte Physik,
Universit\"at Stuttgart, Pfaffenwaldring~57, D-70569 Stuttgart, Germany.}
\affiliation{\small Institute of Physical Chemistry, Polish Academy of
  Sciences, Kasprzaka 44/52, 01-224 Warsaw, Poland.}
\author{S.~Dietrich}
\affiliation{\small Max-Planck-Institut f\"ur Metallforschung,
  Heisenbergstr.~3, 70569 Stuttgart, Germany}
\affiliation{\small Institut f\"ur Theoretische und Angewandte Physik,
Universit\"at Stuttgart, Pfaffenwaldring~57, D-70569 Stuttgart, Germany.}


\begin{abstract}
The confinement of critical fluctuations in soft media
induces critical Casimir forces acting on the confining 
surfaces. The temperature and geometry dependences of such 
forces are characterized by {\it universal} scaling
functions.
A novel approach is presented 
to determine them for films via Monte Carlo simulations of 
lattice models.
The method is based on an integration scheme of free energy
differences. 
Our results for the Ising and the XY universality class compare
favourably with corresponding experimental results for
wetting layers of classical binary liquid mixtures and of $^4$He,
respectively. %
\end{abstract}

\maketitle

\section{Introduction}
Recent progress in understanding the features of 
effective forces induced by confined 
fluctuations, both quantum and thermal, 
reveals the potential  relevance of these
so-called Casimir forces~\cite{Casimir}
for numerous  applications, ranging 
from microelectromechanical systems
(MEMS) to the physics of colloids~\cite{nature,krech:99:0,colloids}.
Thermal fluctuation-induced Casimir forces $f_\Cas$ acting on the confining
surfaces of fluids near critical
points~\cite{FdG} are of particular interest
because they become largely independent of the microscopic details of
the system, acquiring a {\it universal}
character~\cite{FdG,diehl:86:0,krech:92}, and they 
can be switched on and off upon varying, e.g., the temperature.
Moreover, by changing the surface chemistry 
they can be relatively easily turned from attractive to
repulsive~\cite{colloids}, in  contrast  to the Casimir force stemming from 
electromagnetic fluctuations, for which such a possibility is currently
debated as being very desirable to avoid stiction in MEMS, but
difficult to achieve.
Finite-size scaling theory (see, e.g., 
ref.~\cite{krech:99:0}) predicts that the
temperature dependence of the {\it critical}
Casimir force $f_\Cas$ is described by universal scaling functions
%
which depend on the bulk universality class (UC) 
of the confined medium and on the surface UCs of 
the confining surfaces~\cite{diehl:86:0}.
The latter are related to the boundary conditions 
(BC)~\cite{diehl:86:0,krech:99:0} imposed by the surfaces on the relevant
fluctuating field, i.e., the order parameter (OP) of the underlying
second-order phase transition.
In spite of intensive theoretical and experimental efforts, the current
knowledge of these scaling functions is still rather limited even for 
relevant UCs such as the Ising one, 
which characterizes the critical behaviour of
simple fluids and binary liquid mixtures.
In three spatial dimensions (3D) the only  available results
refer, theoretically, to films with periodic BC (PBC), investigated via 
Monte Carlo (MC) simulations~\cite{DK}, or field-theoretical
methods (Dirichlet, Neumann BC, PBC)~\cite{krech:92}, 
and, experimentally, to complete wetting films of 
binary liquid mixture~\cite{pershan} belonging to the surface
UC characterized by symmetry-breaking 
surface fields~\cite{diehl:86:0}.
The corresponding BC ($+-$) of opposing surface fields reflect the fact
that the two confining surfaces exhibit opposite adsorption
preferences for the two species of the mixture. 
{\it At} the {\it bulk} critical point the dependence of $f_\Cas$ 
on the thickness $L$ of the film turned out to be in good agreement with
the corresponding theoretical predictions~\cite{krech,upton}.
However, the determination of the full temperature dependence of
$f_\Cas$ from these very difficult experiments suffers from
significant statistical and systematic uncertainties,
enhancing the need of theoretical insight.
Indeed, several features of the associated scaling function, such as its
global shape and its dependence on the spatial dimensionality $d$ and 
BC still await theoretical investigations.
Exact results are available in $d=2$~\cite{ES} and $d\ge 4$~\cite{krech}
(mean-field theory) both for ($+-$) and ($++$)
BC, the latter corresponding to the case in which both
confining surfaces exhibit preference for the {\it same} species of
the mixture. 
Proposals to measure the temperature dependence of the 
Casimir force between a colloid and a flat surface or between
colloidal particles dissolved in 
a near-critical binary liquid  mixture~\cite{colloids}
call for a detailed theoretical
analysis of the associated scaling behaviour.
The relevant missing pieces, mentioned above, in 
the theoretical analysis of the
scaling behaviour of $f_\Cas$ require to account for the
fluctuations, including the dimensional crossover occurring in a film.
This is a rather challenging task, 
especially if 
the OP profile is inhomogeneous across the film, as in the cases we are
interested in.
With these elements out of reach of current analytical techniques, 
MC simulations provide a useful alternative approach.
The  available MC results for  the $d=3$ Ising model are
restricted to the case of PBC~\cite{DK}, in which the scaling
function of $f_\Cas$ can be determined --- up to a normalization
factor --- by numerically measuring the expectation value 
of a suitable lattice stress tensor.
The purpose of the present contribution is to present a novel approach
for the MC simulation of the Casimir force and to
provide data for the scaling behaviour of $f_\Cas$. We focus on
the Ising UC with the experimentally relevant BC ($++$) and
($+-$). We also compare our results with those in ref.~\cite{DK}, providing
an independent test of the method proposed therein.
Our method is based on an integration scheme of free energy differences
and it has the  advantage, compared to the latter, 
of providing the {\it absolute} value for
$f_\Cas$ and of being applicable for arbitrary BC.
The comparison with the experimental
data in ref.~\cite{pershan} reveals good agreement. 

Measurements~\cite{garcia} of the equilibrium 
thickness of $^4$He wetting films near the superfluid temperature
$T_{\lambda}$ %
provide an experimental determination of the scaling function of
$f_\Cas$ for the XY UC with Dirichlet BC 
on both surfaces, corresponding to the so-called 
ordinary surface UC~\cite{diehl:86:0}. 
These BC are due to the fact that the superfluid OP vanishes at 
the surfaces.
In this case more analytical and numerical results are available.
For temperatures $T\ge T_{\lambda}$ 
field-theoretical calculations~\cite{krech:92} of the scaling function
are in agreement with the experimental data~\cite{garcia}, 
whereas its behaviour for $T\ll T_{\lambda}$ has been 
determined by accounting for He-specific features related to
capillary-wavelike surface fluctuations~\cite{kardar:04}. In addition,
valuable information on the shape of the scaling function in the
critical region has been obtained on the basis of Landau-Ginzburg
theory~\cite{LGW-MF}. Recent MC simulations~\cite{hucht} have nicely
confirmed and extended the available analytic and experimental results.
Since our approach differs from the one in ref.~\cite{hucht}
we also present our results 
for the  scaling function in this case, providing a 
valuable test.


%
In a film geometry with thickness $L$ and large transverse area $A$, 
the Casimir force $f_\Cas$ per unit area $A$ 
and in units of $\kB T \equiv \beta^{-1}$ is defined as
$f_\Cas(\beta,L)\equiv - \partial f^\ex/\partial L$, where
$f^\ex(\beta,L)\equiv \beta L [f-f^\bulk(\beta)]$
is the excess free energy (which depends on the BC), $f$ 
is the free energy of the film per unit volume $V=LA$ %
and  $f^\bulk$ is the bulk free energy density.
According to finite-size scaling~\cite{krech:99:0}
the Casimir force takes the universal scaling form   
\be
\label{eq:scf}
f_\Cas(\beta,L)=L^{-d}\vartheta\left(\tau(L/\xi_0^+)^{1/\nu}\right)
\ee
where the scaling function $\vartheta(x)$ depends on the BC,
$\tau= 
(\beta_{c}-\beta)/\beta$ is the reduced 
temperature and $\xi=\xi_{0}^{\pm} |\tau|^{-\nu}$ is the {\it bulk}
correlation length which controls the spatial 
exponential decay of correlations. 
The critical exponent $\nu$ equals
$0.6301(4)$ and $0.662(7)$ for the 3D Ising and XY
UCs, respectively~\cite{PV}; $\xi_0^\pm$ are
nonuniversal amplitudes above $(+)$ and below $(-)$ $T_c$.
%
%
%
%
\section{Computation of the scaling functions}
We compute the Casimir force for the 
Ising and XY models 
defined on  a 3D simple cubic lattice in slab
geometry ($L_{x} \times L_{y} \times L_{z}$ with $L_{x}=L_{y} \gg 
L_{z} \equiv L$ and $A=L_x \times L_y$)
via the Hamiltonian 
$H = - J \sum_{\la i,j \ra} {\bf s}_{i} \cdot {\bf s}_{j}$, 
where the sum $\la i,j \ra$ is taken over all nearest neighbour pairs
of sites $i$ and $j$ on the lattice. In the Ising model,
${\bf s}_i$ has only one component $s_i \in \{ +1, -1\}$, whereas in
the XY model ${\bf s}_i$ is a two-component vector with modulus 
$|{\bf s}_i| = 1$.
With the Hamiltonian $H$ one finds 
$\beta_c=0.2216544(3)$ 
and $\xi^+_0=0.501(2)$~\cite{RZW}
for the Ising model, whereas $\beta_c=0.45420(2)$ and
$\xi_0^+=0.498(2)$~\cite{GH}, for the XY 
model.\footnote{Note that although the values of $\xi_0^+$ quoted here
refer to the {\it second moment} correlation length $\xi_{\mathrm{2^{nd}}}$,
$\xi/\xi_{\mathrm{2^{nd}}} \simeq 1$ for $\beta < \beta_c$ for both the
Ising and the XY model~\cite{PV,GH}.}
Temperatures and energies 
are measured in units of $J$ and $\xi_0^+$ in units of the lattice 
spacing.
In the $x$ and $y$ directions we assume PBC
whereas in the $z$ direction we consider periodic, free, and fixed BC 
(i.e., for the Ising model, $s_i=+1$ ($+$) or $s_i=-1$ ($-$) at the
boundaries).
For large $A$, the total free energy
$F(\beta,L,A)$ of such systems decomposes as
$F(\beta,L,A) \equiv A L f 
= A [L f^\bulk(\beta) + \beta^{-1} f^\ex(\beta,L)]$,
where $f^\ex(\beta,L=\infty)$ is the contribution to $F$
due to the two isolated surfaces, macroscopically far apart from each
other, whereas $f^\ex(\beta,L) -
f^\ex(\beta,\infty)$ is the $L$-dependent finite-size contribution 
we are interested in.
On the lattice ($\,\hat{}\,$), $f_\Cas(\beta,L)$ is given by 
\be 
\label{eq:force}
\hat f_\Cas(\beta,L-\frac{1}{2},A)\equiv - \frac{\beta \Delta F(\beta,L,A)}{A}
+ \beta f^\bulk(\beta)\,,
\ee
where $\Delta F(\beta,L,A)= F(\beta,L,A)-
F(\beta,L-1,A)$.
%
%

%
%
In general MC methods do note lend themselves to the 
efficient computation of  quantities such as $F$, which cannot be
expressed as suitable ensemble averages. 
However, the free energy difference 
$\Delta F(\beta,L,A)$
we are interested in can be cast in such a form via the
so-called ``coupling parameter approach'' 
(see, e.g., ref.~\cite{Mon}). This is a viable alternative to the method
used in ref.~\cite{DK} 
in which $\Delta F$ has been expressed as the ensemble 
average of a lattice stress tensor, which so far is only applicable
for PBC.
We consider two lattice models with the same configuration space
$\C$ but different Hamiltonians $H_1$ and $H_2$, so that their free 
energies are 
given by $F_i = -\frac{1}{\beta}\ln \sum_{\C} \exp(-\beta H_i)$ where
$\sum_{\C}$ indicates the sum over all possible configurations
belonging to $\C$. 
$F_2-F_1$ can
be conveniently computed by introducing the crossover Hamiltonian
\be
H_{\rm cr}(\lambda)=(1-\lambda)H_1 +\lambda H_2
\label{Hcr}
\ee
which depends on the coupling parameter
$\lambda\in[0,1]$ and interpolates between $H_1$ and $H_2$ as
$\lambda$ increases from 0 to 1. Accordingly, the free
energy $F_{\rm cr}(\lambda) =  -\frac{1}{\beta}\ln \sum_{\C} \exp(-\beta
H_{\rm cr}(\lambda))$ of a system with Hamiltonian $H_{\rm
cr}(\lambda)$ and configuration space $\C$ interpolates between $F_1$
and $F_2$. The derivative of $F_{\rm cr}(\lambda)$ with respect to the
coupling parameter,
\be
\frac{\rmd F_{\rm cr}(\lambda)}{\rmd \lambda} =
\frac{\sum_\C (H_2-H_1) \rme^{-\beta H_{\rm cr}(\lambda)}}{\sum_\C \rme^{-\beta H_{\rm cr}(\lambda)} }
=\la\Delta H\ra_{H_{\rm cr}(\lambda)}\,,
\ee
takes the form of the canonical ensemble average $\la \ldots \ra_{H_{\rm cr}(\lambda)}$ (with Hamiltonian
$H_{\rm cr}(\lambda)$) of $\Delta H \equiv H_2 - H_1$ and therefore it can be
efficiently computed via MC simulations. 
A straightforward
integration over $\lambda$ yields the expression for 
the free energy
difference 
\be
F_2 - F_1 = \int_0^1\!\!\rmd\lambda \, \la\Delta
H\ra_{H_{\rm cr}(\lambda)} \equiv I
\label{DF}
\ee
in terms of an ensemble average (see, e.g., ref.~\cite{Mon}).

\begin{figure}[h]
\centering
\includegraphics[scale=0.468]{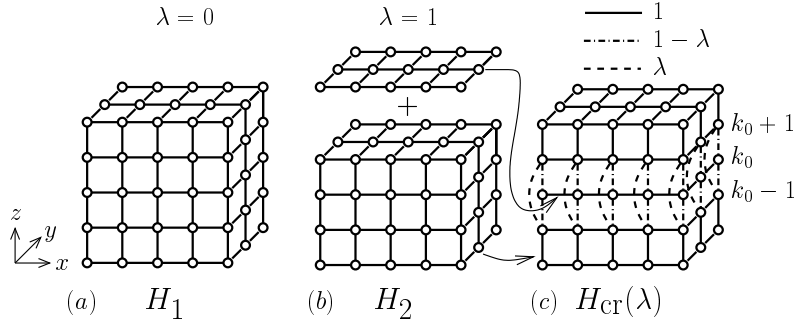}
\caption{%
Bond arrangement for the computation of the free energy 
difference in eq.~\protect{\reff{DF}} (see main text). 
}
\label{fig:fig1}
\end{figure}

The Casimir force 
is related to the difference
$\Delta F(\beta,L,A)$ (see eq.~\reff{eq:force}) 
between the free energies $F(\beta,L,A)$ and $F(\beta,L-1,A)$ of the
same model on two lattices with different numbers of sites and
therefore different configuration spaces. 
In order to apply the method described above for the computation of
$\Delta F(\beta,L,A)$ one identifies the initial Hamiltonian $H_1$
and the configuration space $\C$ with the corresponding ones of the
model on the lattice $A\times L$, as
depicted in fig.~\ref{fig:fig1}(a). Accordingly, 
$F_1(\beta,L,A) = F(\beta,L,A)$.
The configuration space of the final system can be arranged to be equal to
$\C$ by adding to the model on the lattice $A \times (L-1)$
we are actually interested in a two-dimensional lattice of size
$A$ with suitable degrees of freedom and lateral PBC 
(see fig.~\ref{fig:fig1}(b)).
The final Hamiltonian $H_2$ is then constructed such that the
added layer does not interact with the remaining part of the system 
and therefore $F_2(\beta,L,A) = F(\beta,L-1,A) + F_{2D}(\beta,A)$, where
$F_{2D}(\beta,A)$ is the free energy of the isolated
two-dimensional layer. 
Although the argument is quite general, we focus now on the case of
interest in which the lattice degrees of freedom (e.g., spins of the
Ising model) interact only with their nearest neighbours on the same
lattice, with a coupling strength $J=1$ (indicated by solid bonds in
figs.~\ref{fig:fig1} (a) and (b)). 
The crossover Hamiltonian $H_{\rm cr}(\lambda)$ (see eq.~\reff{Hcr})
additionally depends on the position $k_0\in\{1,2,\ldots,L\}$ 
(along the $z$-direction) of the two-dimensional layer which 
decouples
from the rest of the system upon passing from $\lambda=0$ to
$\lambda=1$, i.e., from fig.~\ref{fig:fig1} (a) to (b). 
The resulting $H_{\rm cr}(\lambda)$ is characterized by the coupling
constants depicted in fig.~\ref{fig:fig1}(c) whereas $\Delta H$ (see
eq.~\reff{DF}) can be determined as $\Delta H = H_{\rm cr}(\lambda=1)
- H_{\rm cr}(\lambda=0)$. 
%
$\Delta F$ (see
eqs.~\reff{eq:force} and~\reff{DF}) can be finally expressed as
$\Delta F(\beta,L,A) = - I(\beta,L,A) + F_{2D}(\beta,A)$
from which one has still to subtract 
$f^\bulk(\beta)$ in
order to determine the Casimir force in a
slab of thickness $L-1/2$ (see eq.~\reff{eq:force}).
However, it is numerically more convenient to avoid the computation
of $f^\bulk(\beta)$ 
by considering, instead, the {\it difference} between the Casimir forces
in slabs of thicknesses $L_1$ and $L_2>L_1$:
\be
\begin{split}
\Delta \hat f_\Cas(\beta,L_1,L_2,A) & \equiv 
\hat f_\Cas(\beta,L_{1}-\frac{1}{2},A)-\hat
f_\Cas(\beta,L_{2}-\frac{1}{2},A)\\
&= \beta A^{-1} [I(\beta,L_1,A)-I(\beta,L_2,A)],
\end{split}
\label{eq:deltaC}
\ee 
in which the contributions of both 
$f^\bulk(\beta)$ and $F_{2D}(\beta,A)$ actually 
cancel. 
Below we describe the method used to determine 
$\vartheta$ in eq.~\reff{eq:scf} on the basis
of the numerical data for $\Delta \hat f_\Cas(\beta,L_1,L_2,A)$.
In passing, we note that although $H_{\rm cr}(\lambda)$
(see fig.~\ref{fig:fig1}(c)), $\Delta H$, %
and therefore $\la \Delta H\ra_{H_{\rm cr}(\lambda)}$ depend on the
choice of $k_0$, $\int_0^1\rmd\lambda\,\la \Delta H\ra_{H_{\rm
cr}(\lambda)}$ is actually independent of it, as long as the boundary
conditions are not affected by the extraction of the $k_0$-th layer.
In particular, imposing the BC at the boundary layers in the
$z$-direction, this requires $k_0\neq 1, L$ for fixed and open
BC, whereas for PBC there is
no restriction and indeed translational invariance implies that  $\la \Delta
H\ra_{H_{\rm cr}(\lambda)}$ is actually independent of $k_0$.
In our simulations we have taken $k_0 = L/2$. 
%
%
%

%
%
Within the MC simulations we compute the ensemble averages
$\la \Delta H\ra_{H_{\rm cr}(\lambda)}$ 
for different
values of $\beta$, lattice sizes, and $\lambda$. Then, via numerical
integration (Simpsons method with 20 points) we 
calculate the integral $I(\beta,L,A)$ in eq.~\reff{DF}
and thus
$\Delta \hat f_\Cas(\beta,L_1,L_2,A)$ (see eq.~\reff{eq:deltaC}) 
with $(L_1,L_2) = (L,2 L)$ and $L=13,16,20$ for the Ising
model, whereas $L=10,15,20$ for the XY model. 
For a given pair of thicknesses $(L_1,L_2)$, fixed $A$ and BC, 
the scaling function $\sf$ of the Casimir force can be extracted
from the temperature dependence of $\Delta \hat f_\Cas$
by using the fact that, for large $L_{1,2}$ and $A$, $\hat f_\Cas$ in
eq.~\reff{eq:deltaC} scales according to eq.~\reff{eq:scf}. 
In particular, it is useful to 
focus on the quantity
\be
g(y;L_1,L_2,A)\equiv \left(L_{1}-1/2 \right)^{d} 
\Delta \hat f_\Cas(\beta(y;L_1),L_1,L_2,A)\,,
\ee
as a function of $y$, where $\beta(y;L_1) \equiv  \beta_c/[1+ y
(L_1-1/2)^{-1/\nu}]$
and $d=3$; $g$ is expected to
scale as (see eq.~\reff{eq:scf})
\be
g(y;L_1,L_2,A) = \sfhb(y) - \alpha^{-d} \sfhb(\alpha^{1/\nu} y)\,,
\label{eq:implicit}
\ee
where $\alpha = (L_2-1/2)/(L_1-1/2)$ is the width ratio and $\sfhb(y)$
is the lattice estimate of $\sfb(y) \equiv \sf(y/(\xi_0^+)^{1/\nu})$. 
For given $g$ eq.~\reff{eq:implicit} can be solved for $\sfhb(y)$ 
via an iterative
method. In the first step one takes 
$\sfhb_{0}(y)
\equiv g(y;L_{1},L_{2},A)$ as a first approximation of the actual 
$\sfhb$. In turn, this approximant
can be improved by taking into account that eq.~\reff{eq:implicit} yields
$\sfhb(y) = \sfhb_0(y) + \alpha^{-d}\sfhb(\alpha^{1/\nu}y) \simeq
\sfhb_0(y) + \alpha^{-d}\sfhb_0(\alpha^{1/\nu}y)$, so that a better 
approximant $\sfhb_1(y)$ is provided by
$\sfhb_1(y) = \sfhb_0(y) + \alpha^{-d}\sfhb_0(\alpha^{1/\nu}y)$ 
where the value of $\sfh_0$ at the 
point $\alpha^{1/\nu} y$ is obtained by cubic spline interpolation of
the available data. 
In the expression for $\sfhb_1$ one can replace $\sfhb_0$ by using
eq.~\reff{eq:implicit}, yielding 
$\sfhb_{1}(y)=\sfhb(y)-\alpha^{- 2d}\sfhb(\alpha^{2/\nu} y)$, which, in
turn, can be solved as already done before for eq.~\reff{eq:implicit} by
introducing $\sfhb_{2}(y)=\sfhb_{1}(y)+\alpha^{-2 d}\sfhb_{1}( \alpha^{2/
\nu} y) = \sfhb(y)-\alpha^{- 4d}\sfhb(\alpha^{4/ \nu} y)$, and so
on. This iterative procedure yields a sequence of approximants
$\sfhb_{k\ge1}(y)=\sfhb_{k-1}(y)+\alpha^{-2^{k-1} d}\sfhb_{k-1}( 
\alpha^{2^{k-1}/\nu} y)$,
which converges very rapidly because the correction
to the $k$-th approximant is 
of the order of $\alpha^{-2^{k-1}d}$, i.e., exponentially small 
in $2^k$.
%
Already for $k=5$ one has $\alpha^{-2^{k-1}d}\simeq 3.5 \times 10^{-15}$ in
3D  with $\alpha \simeq 2$(\footnote{%
Note that a smaller value of $\alpha$ reduces the sizes of the
lattices required for the computation of $\hat f_\Cas$. On the other
hand, such a choice decreases the accuracy of the iterative procedure 
to determine $\hat \sfb$.%
}).
Accordingly, we approximate $\sfhb(y) \equiv \sfhb_{k\to\infty}(y)$ by
$\sfhb_5(y)$. 
%
%
%
The result for the universal scaling function 
$\sfh(x) \equiv \sfhb(x (\xi_0^+)^{1/\nu})$ of $\hat f_\Cas$, 
as obtained
from a specific pair of lattices with thicknesses $(L,2L)$, should be
independent of the actual value of $L$, at least for large $L$. 
However, for the thicknesses we used in our MC simulations,
corrections to the leading scaling behaviour are actually
relevant~\cite{hucht} and affect both the scaling variable 
\be
\label{eq:xcorr}
x \equiv \tau (L/\xi_0^+)^{1/\nu} (1 + g_\omega L^{-\omega})
\ee
and the scaling function $\sfh(x)$ 
which additionally depends on $L^{-\omega'}$: $\hat f_\Cas(\beta,L,A)
= L^{-d} \sfh(x,L^{-\omega'}) \simeq L^{-d} \sf(x) [1 +
L^{-\omega'} \phi(x)  + \ldots]$ for large $L$. In
eq.~\reff{eq:xcorr}, $\omega$ is the leading correction-to-scaling
exponent $\omega \simeq 0.84(4)$ and $0.79$~\cite{PV}
for the Ising and XY UC, respectively, in 3D. 
$\omega'$ controls the leading corrections
to scaling of the estimator $\hat f_\Cas$ and its value, typically
equal to $\omega$, can be increased by using suitably improved
Hamiltonians and observables~\cite{PV} so that these corrections are 
reduced. 
However, 
we point out that in the present case surface operators~\cite{diehl:86:0} 
might even yield $\omega'<\omega$. For small lattice sizes,
next-to-leading corrections to scaling (e.g., $\sim L^{-1}$) might also be of
relevance, resulting in effective $L$-dependent exponents.
A detailed analysis of all these corrections and the
determination of $\phi(x)$ and $\omega'$ is beyond the scope of the
present paper and requires the study of much larger lattices. 
As a phenomenological ansatz for the effective corrections we take 
$\omega' \simeq \omega \simeq 1$ and~\cite{hucht}
\be
\label{eq:sfcorr}
\hat f_\Cas(\beta,L,A) = L^{-d} (1 + g_1 L^{-1})^{-1} \sf(x) \,,
\ee
which, for suitable choices of $g_1$ and
$g_\omega$, yields a good data collapse of the curves
corresponding to different sizes.
We point out that an equally satisfactory data collapse can
be obtained --- within the range of the scaling variable $x$ explored here --- 
by assuming a different functional form for the corrections. The
resulting estimate of $\sf(x)$ is slightly affected by this choice and
only larger scale simulations can provide an  estimate of
$\sf(x)$ which is unbiased in this respect.
In addition to these corrections to scaling, the simulation data 
depend on the aspect ratio $\rho \equiv L/\sqrt{A}$. Whereas in the
case of the XY model this dependence 
is quite pronounced~\cite{hucht}, for the Ising model with ($++$)
and ($+-$) BC and for $x\ge -6$, 
$\Delta \hat f_\Cas(\beta, L, 2 L,A)$ exhibits only a very weak dependence
on $\rho$ already for $\rho \le 1/6$, as we have tested by considering
lattices with $1/\rho = 6, 10, 14$.
The results we present here for the Ising model refer to lattices with 
fixed $\rho=1/6$.
For the XY model we have accounted for corrections due to $\rho\neq 0$
in accordance with ref.~\cite{hucht} by considering multiplicative
corrections $1+r_1\rho^2$ and $(1+r_2\rho^2)^{-1}$ 
to $x$ and $\hat f_\Cas$, respectively, 
which allow the extrapolation of the data 
on lattices with $1/\rho=4, 6, 8$ to $\rho\to 0$.

%
%
%
The computation of the canonical average 
$\la \Delta H\ra_{H_{\rm cr}(\lambda)}$ has been carried out via an 
hybrid MC method in which a suitable mixture of Wolff 
and  Metropolis algorithms is used~\cite{LB}.
In particular, for the Ising model each hybrid MC step consists of 
one flip of a Wolff cluster according to the Wolff algorithm, 
typically followed by $9 A$  
attempts to flip a spin $s_{x,y,z}$ with $z\in \{k_0-1,k_0,k_0+1\}$,
which are accepted according to the Metropolis rate~\cite{LB}. An analogous
method, with a suitable implementation of Metropolis and Wolff algorithms has
been used for the XY model~\cite{GH}.
%
%
%
We tested our MC program successfully by comparing $g(y,5,10,9)$ with
corresponding transfer-matrix data.
%
%
%

\section{3D XY model} 
\begin{figure}[h]
\centering
\includegraphics[scale=0.78]{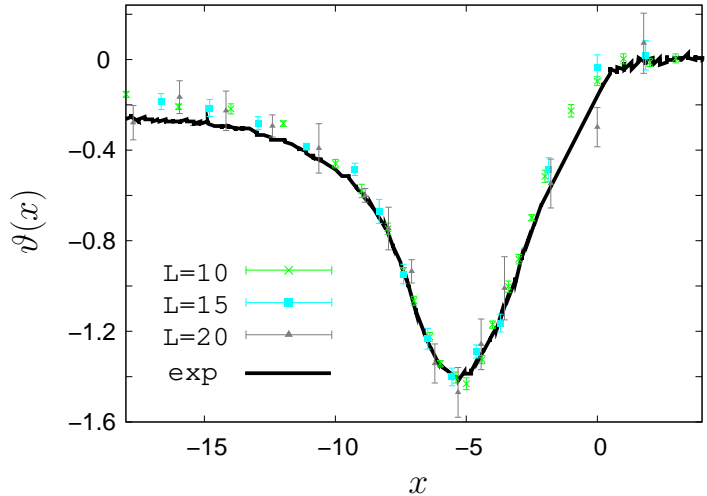}
\caption{Scaling function $\sf$ of the Casimir force for the
3D XY bulk UC and so-called ordinary surface
UC corresponding to free boundary conditions. Our MC
data compare very well with the corresponding experimental data
from ref.~\cite{garcia} (solid line).}
\label{fig:xy_scaling}
\end{figure}

%
We have determined the scaling function of the Casimir
force in the 3D XY model with free BC, 
which is of relevance for the $^4$He experiment mentioned in the 
introduction.
The resulting scaling function is plotted in
fig.~\ref{fig:xy_scaling}. In order to achieve scaling we have
accounted for corrections to scaling according to
eq.~\reff{eq:sfcorr} with 
$g_1 = 6.4(2)$, $g_\omega = 2.1(2)$ 
and for
the corrections due to $\rho\neq 0$ in accordance with ref.~\cite{hucht} 
with $r_1 = 2.3(2)$ and $r_2=1.1(1)$  %
(see ref.~\cite{hucht} for details). The scaling function in
fig.~\ref{fig:xy_scaling} is compatible, within the errorbars, with
the one determined in ref.~\cite{hucht}, providing an independent test both
of the results presented there and of our method to compute it. 
%
In the approach used in ref.~\cite{hucht}, 
$\hat f_\Cas$ is computed via the internal energy density $u$ and an
integration over the temperature, whereas 
here this is carried out via the free energy density $f$ and 
an integration over the coupling $\lambda$.
In ref.~\cite{hucht} one takes advantage of the possibly
available numerical knowledge of the bulk energy density
$u^\bulk$ of the model of interest whereas here the analogous information
on $f^\bulk$ is not required for the determination of
$\hat\sf$, making our approach applicable also to cases in which there
is no detailed knowledge of $u^\bulk$ and $f^\bulk$.
%
%
%
%
%
The MC results for $\vartheta(x)$ in fig.~\ref{fig:xy_scaling} 
compare also very well with the experimental data 
in ref.~\cite{garcia}
(we have used the experimental value 
$\xi_0^{+\mathrm{(exp)}} = 1.432$\AA\ 
for the normalization of $x$). 
In particular 
with $x_\mathrm{min} = -5.29(7)$
and $\sf_\mathrm{min} \equiv \sf(x_{\mathrm{min}}) = -1.41(2)$, 
it captures properly the corresponding experimental values
$x_{\mathrm{min}}^\mathrm{(exp)} = -5.7(5)$ and
$\sf_\mathrm{min}^\mathrm{(exp)} = -1.30(3)$ 
for the pronounced minimum. 
In comparing $\sf_\mathrm{min}$ with $\sf_\mathrm{min}^\mathrm{(exp)}$ one has
to take into account that, 
as pointed out above, the numerical determination of $\sf$ is
actually influenced by the choice of the ansatz for the corrections to
scaling. Indeed, replacing the multiplicative correction
$(1+g_1L^{-1})^{-1}$ in eq.~\reff{eq:sfcorr} with one of the form 
$(1-\tilde g_1L^{-1})$ (which is equivalent to the previous one for
large $L$), the resulting scaling function 
displays a good data
collapse for $\tilde g_1 = 3.1(1)$. It has the same
shape as the one in fig.~\ref{fig:xy_scaling} but its overall
amplitude is reduced by a factor $R \simeq 0.89$. %
With this caveat, our estimates for
$x_\mathrm{min}$ and $\sf_\mathrm{min}$ are compatible also with 
those of ref.~\cite{hucht} ($-5.3(1)$ and $-1.35(3)$, respectively).
%
%
%
%
%
%

\section{3D Ising model}
\begin{figure}[h]
\centering
\includegraphics[scale=0.78]{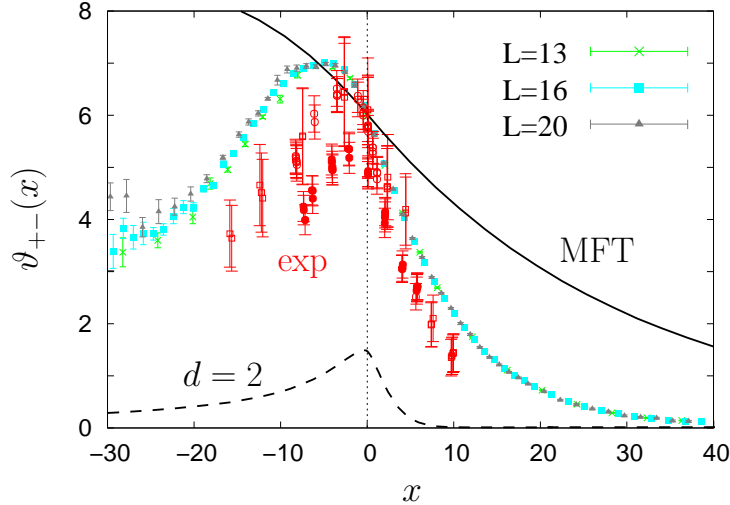}
\caption{Scaling function $\sf_{+-}$ 
of the Casimir force in the 3D Ising model with ($+-$) boundary
conditions, compared with the mean-field prediction (solid line), the
experimental data of ref.~\cite{pershan} and the exact result for the
two-dimensional Ising model (dashed line).}
\label{fig:Ising_pm_scaling}
\end{figure}

\begin{figure}[h]
\centering
\includegraphics[scale=0.78]{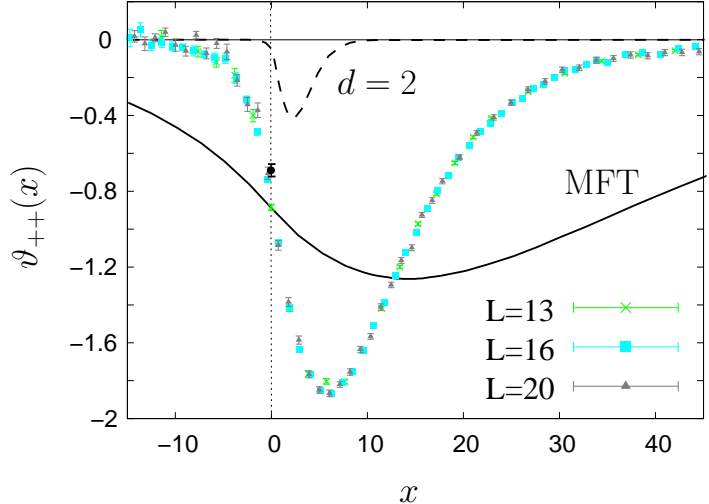}
\caption{Scaling function $\sf_{++}$ of the Casimir force in the 3D Ising model
  with ($++$) boundary conditions, compared with the mean-field
prediction (solid line)  and the exact result for the
two-dimensional Ising model (dashed line); 
$\bullet$ MC~\protect{\cite{krech}}.}
\label{fig:Ising_pp_scaling}
\end{figure}

\begin{figure}[h]
\centering
\includegraphics[scale=0.504]{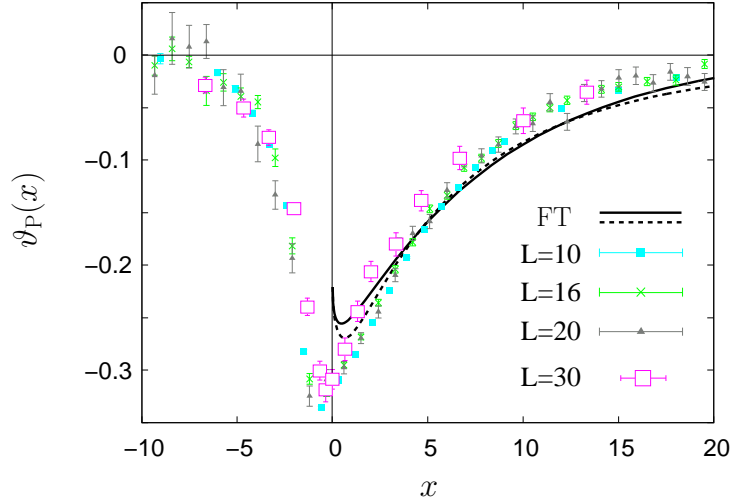}
\caption{Scaling function $\sf_{\mathrm{P}}$ 
of the Casimir force in the 3D Ising model
  with PBC. 
For comparison we report the normalized data
set with $L=30$ from ref.~\protect{\cite{DK}} which corresponds to the
largest lattice size investigated therein. The solid (dashed)
curve corresponds to the [1,0] ([0,1]) Pad\'e approximant of the
analytical prediction in refs.~\cite{krech:92,krech:99:0}.
}
\label{fig:Ising_pbc_scaling}
\end{figure}

For this bulk UC 
we have determined the scaling functions $\sf$ for three 
different BC: ($+-$), 
($++$) (pairs of lattices with sizes $(L,2L)$ and $L=13,16,20$) and PBC 
($L=10,16,20$).  The results are reported in figs.~\ref{fig:Ising_pm_scaling}, \ref{fig:Ising_pp_scaling}
and~\ref{fig:Ising_pbc_scaling}, respectively. 
Each data point has been averaged over at least $10^5$
hybrid MC steps.
The scaling function in fig.~\ref{fig:Ising_pm_scaling} has been
obtained accounting for the corrections according to
eq.~\reff{eq:sfcorr} with 
$g_1 = 14.8(2)$ and $g_\omega = 2.9(2)$. 
The resulting data collapse is very good and only at very low
temperatures corrections to scaling are stronger and not fully
accounted for by the ansatz in eq.~\reff{eq:sfcorr}. 
Such stronger corrections might be
related to the presence of an interface in the system. 
$\sf_{+-}$ is compared with the experimental data
in ref.~\cite{pershan}, the prediction of
mean-field theory~\cite{krech} (solid line, normalized such that 
$\sf_{+-}^\mathrm{(MFT)}(0) = \sf_{+-}^\mathrm{(MC)}(0)$) 
and the corresponding result for the two-dimensional Ising
model~\cite{ES} (dashed line). %
From the data set with $L=13$ we estimate $\sf_{+-}(0) = 5.97(2)$, in
agreement with the experimental value $\sf^\mathrm{(exp)}_{+-}(0)
=6(2)$~\cite{pershan} but 
which is larger compared to the previous MC estimate $\sf_{+-}(0) =
4.900(64)$~\cite{krech} and the analytical estimates
$\sf_{+-}^\mathrm{(FT)}(0) = 3.16, 4.78$. The latter depend on the
approximant used to resum the field-theoretical $\epsilon = 4-d$-expansion
up to $O(\epsilon)$ (see ref.~\cite{krech} for details). 
A good data collapse is obtained also for $\tilde g_1 =  5.0(1)$ which, 
compared to 
fig.~\ref{fig:Ising_pm_scaling},
yields an overall reduction of the amplitude of $\sf_{+-}$ 
by a factor $R\simeq 0.76$.
In addition,
we expect the experimental data in ref.~\cite{pershan} to
be affected by corrections to scaling already for $x\gtrsim 2$, 
due to the relatively small corresponding value of
$\xi/\ell \lesssim 30$, where $\ell \simeq 3$\AA\ is the molecular
scale in the experiment.
In view of these
difficulties, the agreement between the MC and the experimental data 
in fig.~\ref{fig:Ising_pm_scaling} is encouraging.
%
%
%
In fig.~\ref{fig:Ising_pp_scaling} 
the scaling function $\sf_{++}$ for ($++$) BC has been
obtained accounting for the corrections in eq.~\reff{eq:sfcorr} with 
$g_1 = 14.2(7)$ and $g_\omega = 2.3(2)$. 
Using, instead, $\tilde g_1 = 4.9(2)$ yields $R \simeq 0.77$.
Currently, for this BC and in film geometry no experimental data 
are available for comparison, but $\sf_{++}$ can be compared with the
prediction of mean-field theory~\cite{krech} (solid line, normalized as
before) and of the two-dimensional Ising model~\cite{ES} (dashed line). 
From the data with $L=13$ we  estimate
$\sf_{++}(0) = -0.884(16)$ which is slightly larger than the previous MC
estimate $\sf_{++}(0) = -0.690(32)$~\cite{krech} 
(indicated as a black dot in
fig.~\ref{fig:Ising_pp_scaling}, still affected by
finite-size corrections) and the corresponding field-theoretical
predictions $\sf_{++}^\mathrm{(FT)}(0) = -0.652, -0.346$ depending on
the approximant used to resum the $O(\epsilon)$ series 
(see ref.~\cite{krech} for details).
%

%
%
For PBC and $L \ge 10$ corrections to scaling turn out to be 
negligible and 
the resulting scaling function  
$\sf_{\mathrm{P}}$ 
in fig.~\ref{fig:Ising_pbc_scaling}
is in very good agreement 
with its previous determination in ref.~\cite{DK} 
based on the computation of the lattice stress tensor. 
The slight discrepancies might be due to the uncertainty in the 
normalization factor which had to be used in ref.~\cite{DK}.
This agreement provides additional support concerning the reliability
of our approach. 
Figure~\ref{fig:Ising_pbc_scaling} shows also the comparison with the
available field-theoretical predictions~\cite{krech:92} (solid and
dashed lines) for $x\ge 0$ up to $O(\epsilon)$. The 
discrepancies can be traced back to higher-order nonanalytic 
contributions $\sim \epsilon^{3/2}$~\cite{DGS-06}. 
%
%
%
%
\section{Conclusions}
We have presented a novel general approach to determine the universal 
scaling functions $\sf$ of Casimir forces via Monte Carlo simulations.  
The corresponding 
results for the three-dimensional
Ising and XY UCs compare favourably with
previous experimental, numerical and analytical results.
Our predictions for $\sf_{++}$ and $\sf_{+-}$ in a film geometry 
might also contribute to the understanding of 
Casimir forces acting on colloidal particles.

%

\acknowledgments
The authors acknowledge the important contribution of M. De Prato
to the early stages of this work.
They are grateful to A. Hucht, M. Krech and E. Vicari for useful
discussions and to  M. Fukuto and R. Garcia for providing the
experimental data.

%

\end{document}